\documentstyle[12pt,aasms4]{article}
\lefthead{Igumenshchev et al.}
\righthead{Hard X-ray emitting black hole}

\begin{document}

\title{Hard X-ray emitting black hole fed by accretion of \\
       low angular momentum matter}

\author{Igor V. Igumenshchev}
\affil{Institute of Astronomy, 48 Pyatnitskaya Street, 109017 Moscow, Russia;
       \\ ivi@fy.chalmers.se}

\author{Andrei F. Illarionov}
\affil{Astro Space Center of P. N. Lebedev Physical Institute,
       84/32 Profsoyuznaya Street, \\ 117810 Moscow, Russia;
       illarion@lukash.asc.rssi.ru}

\and

\author{Marek A. Abramowicz\altaffilmark{1,2}}
\affil{Institute of Theoretical Physics, G{\"o}teborg University and 
       Chalmers University \\ of Technology, 412 96 G{\"o}teborg, Sweden;
       marek@fy.chalmers.se}

\altaffiltext{1}{Nordita, Blegdamsvej 17, DK-2100 Copenhagen \O, Denmark}
\altaffiltext{2}{Laboratorio Interdisciplinare SISSA, Trieste, Italy, and ICTP,
           Trieste, Italy}

\begin{abstract}
Observed spectra of Active Galactic Nuclei (AGN) and 
luminous X-ray binaries in our Galaxy
suggest that both hot ($\sim 10^9 K$) and cold ($\sim 10^6 K$) plasma 
components exist close to the central accreting black hole.
Hard X-ray component of the spectra is usually explained by Compton
upscattering of optical/UV photons from optically thick cold plasma 
by hot electrons. 
Observations also indicate that some of these objects are quite efficient in 
converting gravitational energy of accretion matter into radiation.
Existing theoretical models have difficulties
in explaining the two plasma components and high intensity of hard X-rays.
Most of the models assume that the hot component
emerges from the cold one due to some kind of instability, but no one
offers a satisfactory physical explanation for this.
Here we propose a solution
to these difficulties that reverses what was imagined previously: in
our model the hot component forms first and afterward it cools down 
to form the cold component.
In our model, accretion flow has initially a small angular momentum,
and thus it has a quasi-spherical geometry at large radii.
Close to the black hole, the accreting matter is heated up in shocks
that form due to the action of the centrifugal force.
The hot post-shock matter is very efficiently cooled down by Comptonization
of low energy photons and condensates into a thin and cool accretion
disk. The thin disk emits the low energy photons which cool the hot
component.
All the properties of our model, in particular the
existence of hot and cold components, follow from an exact numerical
solution of standard hydrodynamical equations --- we postulate no
unknown processes operating in the flow.
In contrast to the recently discussed ADAF, the particular type of
accretion flow considered in this Letter
is both very hot and quite radiatively efficient.
%and it is very promising in explaining the luminous X-ray emitting
%objects. 
\end{abstract}

\keywords{accretion, accretion disks --- black hole physics ---
hydrodynamics --- methods: numerical --- X-rays:general}

\section{Introduction}

There is a general agreement that the observed properties of galactic
black hole candidates and of active galactic nuclei (AGN) could be best
explained in the framework of accretion disks around black holes.
However, no theoretical accretion disk model could explain all the basic
properties of these sources. In particular, the best known, standard
Shakura \& Sunyaev (1973) disk model (SSD),
predicts a temperature of accreted
matter far too low to explain the hard X-ray emission
($h\nu\ga 10 {\rm ke}V$) that is observed.
The observed hard X-ray emission could be explained
by postulating the existence of a very hot plasma, with electron
temperature $T_e \simeq 10^9$~K, in which soft photons emitted by SSD are
boosted to higher energies by inverse Compton effect. The question is,
how does such a hot plasma form in black hole accretion flows.

The very popular disk-corona (DC) model
(e.g., Liang \& Price 1977; Haardt \& Maraschi 1993) postulates
the existence of the $T_e \simeq 10^9$~K plasma in the
form of a hot corona above the cold disk. Because the physics of DC
models is largely ad hoc, a typical specific DC model contains a set of
free tunable parameters.

In the Shapiro, Lightman \& Eardley (1976, SLE) hot, optically thin
accretion disk model, ions are heated by viscous dissipation of their
orbital energy, and inefficiently cooled by the Coulomb interaction with
electrons. Thus, the ions have temperature close to the virial
temperature, $10^{11}-10^{12}$~K. Electrons are very efficiently cooled
by a variety of radiative mechanisms, and this reduces their temperature
to about $10^9-10^{10}$~K, which is sufficient for explanation of the hard
X-ray radiation. The SLE model is, however, violently thermally unstable
and therefore it cannot describe really existing objects.

Detailed models of black hole optically thin accretion flows in which
cooling is dominated by advection (ADAF) have been recently constructed
in many papers (see recent reviews in
Abramowicz, Bj\"ornsson \& Pringle 1998 and
Kato, Fukue \& Mineshige 1998). ADAFs are
hot, with the electron temperature about $10^9-10^{10}$~K,
and underluminous, $L \ll L_{Edd}$.
Here $L_{Edd}=1.3\times 10^{38} (M/M_{\odot})$ erg s$^{-1}$
is the Eddington luminosity corresponding to the black hole mass $M$.
No ADAF
solution is possible above a limiting accretion rate that is roughly
about $0.1~{\dot M}_{Edd} = 0.1~L_{Edd}/c^2$.
In some black hole sources, however, observations point to accretion rates
and radiative efficiencies that are much too high for
the standard ADAF model to
explain (see review in Szuszkiewicz, Malkan \& Abramowicz 1996).
On theoretical
side, no satisfactory model of an SSD-ADAF transition has been
worked out and thus, in order to account for the presence of cold matter
together with hot ADAF plasma, one still uses phenomenological arguments
(see Kato \& Nakamura 1998).

A model of inhomogeneous inner region of accretion disk was proposed by
Krolik (1998). In this model the accretion flow consists of clouds 
moving in a hot, magnetized intercloud medium,
which can principally explain the emittion of hard X-rays. 
Such a structure could result from
a dynamical photon bubble instability in radiation pressure supported
disks (Spiegel 1977; Gammie 1998). 
Due to a significant complexity in describing
of the inhomogeneous medium, the model contains a number of phenomenological
assumptions.
In particularly, the assumptions of stability of such a configuration
is questionable.

Motivated by the above difficulties of the
existent models for black hole accretion flow
in explaining the co-existence of the hot and cold components, 
we have constructed a
new model that has the following properties.
(1) Both very hot, hard X-ray emitting gas, and
sufficiently cool gas, consistent with the detection of the fluorescent
iron line, are present very close to the central black hole.
(2) Significant part (up to $\sim 50 \%$) of total luminosity
($\sim 10^{-1}- 10^{-2}L_{Edd}$)
of the object is emitted in hard X-rays.
(3) Flow is stationary and globally stable for large accretion rates
($\dot{M}\sim\dot{M}_{Edd}$).

The key element of the model is that accretion matter has initially a low
angular momentum.
We keep in mind two kind of objects, in which a low angular
momentum accretion onto black hole may occur.
First, black holes, which are fed by accretion from a winds blowing of OB star
in binary systems (Illarionov \& Sunyaev 1975a).
The most popular object of this kind is
X-ray binary Cygnus X-1 (see Liang \& Nolan 1984).
Second, luminous X-ray quasars and AGN, where the central
supermassive black hole
is fed by the matter, which is lost from stars of slowly rotating
central stellar cluster (Illarionov 1988).
In its basic respects, the geometry of our model closely
resembles that of
the DC model, 
but there are also significant differences.
Our models is also very different from that recently discussed by
Esin (1997) although they both stress the importance of
cooling by inverse Comptonization.

\section{Low angular momentum accretion}

Let us consider matter 
with a low characteristic specific angular momentum $\bar{\ell}$,
accreted quasi-spherically onto the central black hole.
By low $\bar{\ell}$ we mean that which
corresponds to the Keplerian orbit with radius
$r_0=2\bar{\ell}^2/r_g c^2$ in  range,
$3r_g < r_0 \la 100 r_g$.
Here $r_g = 3\times 10^5 (M/M_\odot)$ cm is the gravitational radius of
black hole with the mass $M$. Matter with angular momentum smaller than
that in the indicated range could not
form an accretion
disk around the Schwarzschild black hole because the innermost
stable orbit around such a hole is localized at $r_{ms} = 3r_g$.
Matter with $\bar{\ell}$
higher than in the indicated range could
form an accretion disk that extends from the black hole to large radial
distances, $r \ga 100 r_g$: this would correspond to the previously studied
SSD.

At large radii, $r \gg r_0$, the low angular momentum
accretion flow closely resembles spherical Bondi accretion
(Bondi 1952).
Approximated models of spherical accretion onto
a luminous X-ray central source was studied by several
authors (e.g. Ostriker et al. 1976; Bisnovatyi-Kogan \& Blinnikov 1980).
It was shown by Igumenshchev, Illarionov \& Kompaneets (1993)
that inside of the
Compton radius $r_C = 10^4(10^8K/T_C)r_g$, where the Compton
temperature $T_C$ is determined by the `average' photon energy
of the source, accretion
is almost spherical and supersonic. 
Our model shows that at
smaller radii $r\sim r_0$, the flow significantly
deviates from the spherical accretion. Fluid elements tend to cross the
equatorial plane at the radius which corresponds to the Keplerian orbit
for the angular momentum of the element. This leads to formation of
shocks above and below the equatorial plane at $r\la r_0$. By
crossing the shocks, protons reach the virial temperature $T_p \approx
10^{12}(r_g/r)\,K$. In the presence of soft photons from the thin
accretion disk, electrons in the post-shock region are efficiently cooled
by inverse Comptonization. These cold electrons also efficiently cool
protons via Coulomb collisions. 
%Due to the drop of proton temperature and
%the increase of plasma density
%(under roughly isobaric condition in the post-shock region) the
%Bremsstrahlung cooling begins to dominate.
Such a plasma undergoes a runaway cooling 
due to the intensive bremsstrahlung-Compton processes at layers,
where the Compton $y$-parameter reaches the order of unity.
At these layers protons loose the most of their thermal energy,
and the plasma condenses into the thin and cold disk.

The thin disk spreads to very large radii, $r\gg r_0$,
due to the viscous diffusion
(von Weizs\"acker 1948, also see Pringle 1981)
from the region of condensation at $r\sim r_0$.
It is convenient to consider two different parts of the thin disk. 
The inner part, $r\la r_0$,
is an accretion disk, where matter moves inward and angular momentum
is transported outward.
Matter mainly enters the black hole through this
accretion disk, which means that radiative efficiency of
the system is as high as in the standard SSD model
($\approx 6\%$ for the Schwarzschild black hole).
At the outer part of the disk, $r\ga r_0$, angular momentum is
transported outward, removing its excess from the accretion flow to
large radii.

\section{Numerical method}

To study details of the model briefly described 
in the previous Section, we have
simulated the low angular momentum accretion 
of weakly magnetized plasma
onto black hole with the
help of two-dimensional time-dependent hydrodynamical calculations. Our
code is based on the PPM hydrodynamical scheme
(Colella \& Woodward 1984), and solves
non-relativistic Navier-Stokes equations in spherical coordinates,
assuming the azimuthal symmetry. We calculate separately the
balance of electrons and protons internal energies 
%in the ideal-fluid approximation 
for the electron-proton plasma.  The energies of electrons
and protons are coupled through the Coulomb collisions.  
Electrons are cooled by bremsstrahlung and Compton mechanisms. 
The cold and thin accretion disk
at the equatorial plane emits soft photons needed for the Compton
cooling. 
The thickness of the disk is not resolved in our models.
We approximately calculate the energy release
in the disk, and the correspondent radiation flux,
which directly affects the efficiency of the Compton cooling,
using the standard SSD model.
%in which the
%accretion rate equals rate, directly calculated in simulations, of
%condensation of the hot cooling plasma into the disk. 
In the model, the accretion rate in the cold disk equals to the
condensation rate of the hot plasma
into the disk, which is directly calculated in numerical simulations.

We neglect the multiple photon scattering processes,
when calculating the Compton cooling of plasma.
This approximation is quite reasonable for the optically thin flows
($\tau\la 1$), which are characteristic for our models.
We do not solve the transfer equation for radiation emitted by
the thin disk. However, we calculate the radiation density
(which is used to find the plasma energy losses) at each point
above and below the disk by neglecting the absorption of photons.

In our simulations,
protons are heated by shocks, adiabatic compression and viscous
dissipation. We take into account all the components of the viscous
stress tensor corresponding to shear in all directions.
The bulk viscosity is not considered.
The kinematic viscosity coefficient is taken in
the standard $\alpha$-parameterization form:  $\nu=\alpha c_s^2/\Omega_K$,
where $\alpha$ is a constant, $c_s$ is the isothermal sound speed, and
$\Omega_K=(r_g/r)^{3/2}c/\sqrt{2}r_g$
is the Keplerian angular velocity.
We use $\alpha=0.1$ in numerical models.
No artificial numerical viscosity was used in the calculations.

At the outer boundary we assume a supersonic matter inflow
with the
spherically symmetric distributed density and the
specific angular momentum distributed in the polar direction consistently
with rigid rotation: $\ell(\theta)=\ell_{max}\sin^2(\theta)$.
The parameter $\ell_{max}$ determines (roughly)  the
maximum radius,
$r_d=2\ell_{max}^2/r_g c^2$, inside which the
condensation of the hot plasma to the thin disk takes place.
For convenience, we will use the parameter $r_d$ rather than $\ell_{max}$,
when describing results of numerical simulations.
At the
inner boundary, $r_{in}=3 r_g$, and at the equatorial plane, where the
condensation occurs, we assume total absorption of the inflowing matter.

\section{Results and discussions}

We follow the evolution from an initial state, until a time
when a stationary flow pattern is established.
Models show a strong dependence on two parameters:
accretion rate $\dot{M}$, and $r_d$.
We have calculated several models with different $\dot{M}$ and $r_d$.
Two examples of stationary accretion flows with
$r_d=30 r_g$ and two different $\dot{M}=0.25$ and $0.5 \dot{M}_{Edd}$
are presented in Fig.~1.
One can clearly see
two stationary shock structures, which are developed in each model.
The structures are more compact for larger $\dot{M}$.
The nature of
inner shock, at radial distances $r<r_d$ 
above and below the equatorial plane,
has already been discussed in Section~2.
In these shocks, the supersonic accretion flow is slowed
down before condensation to the thin disk. We expect that, in
the hot post-shock plasma, the inverse Comptonization of soft photons
emitted from the thin disk provides the main contribution to the hard X-ray
luminosity of this type of accretion flow. In this region the
maximum proton temperature $T_p\simeq 10^{11}K$ and the electron
temperature $T_e\simeq 10^9 K$ in both models shown in Fig.~1.
The distribution of $T_e$ in the post-shock region is quite uniform. 

\begin{figure}
\plottwo{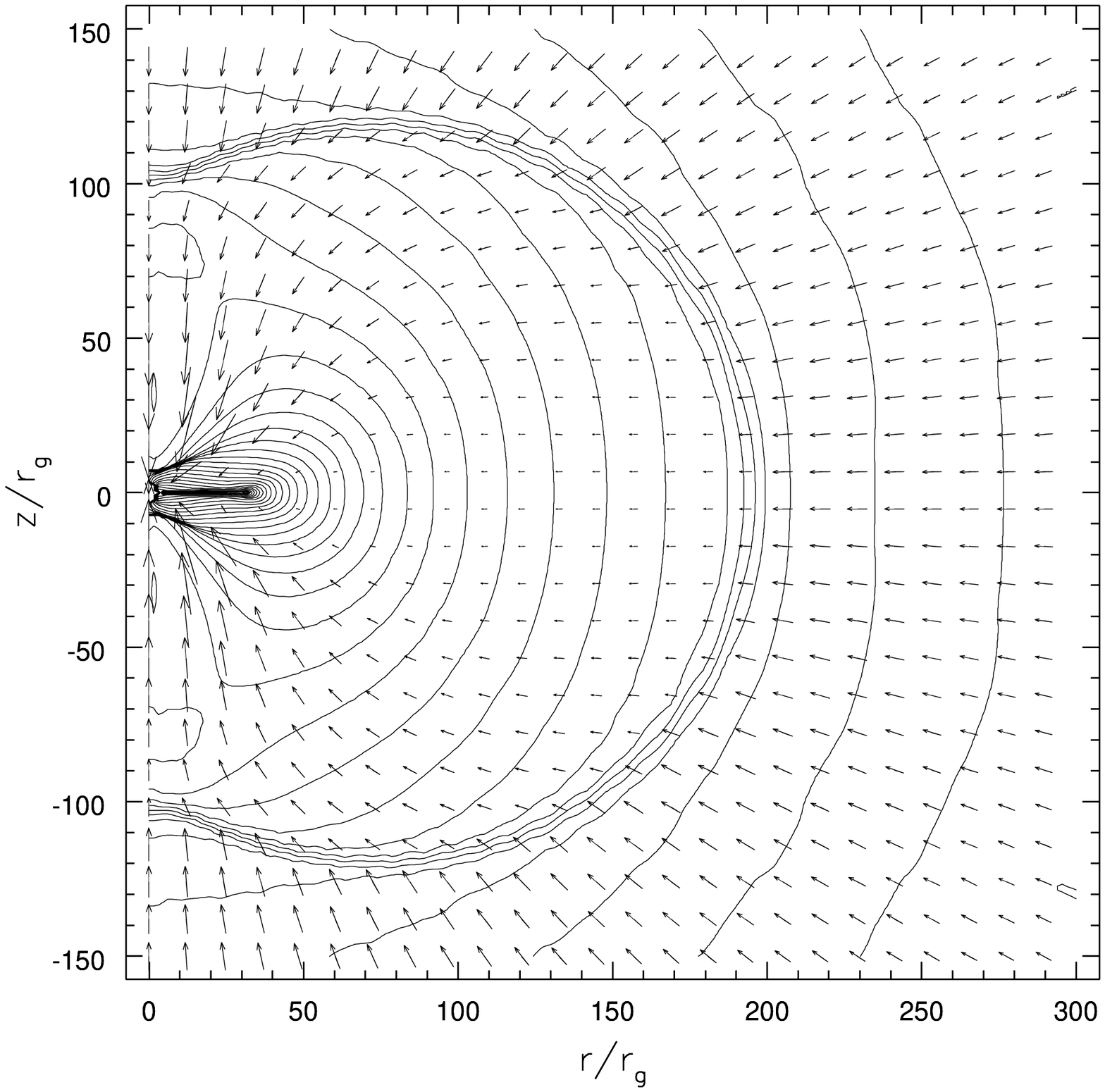}{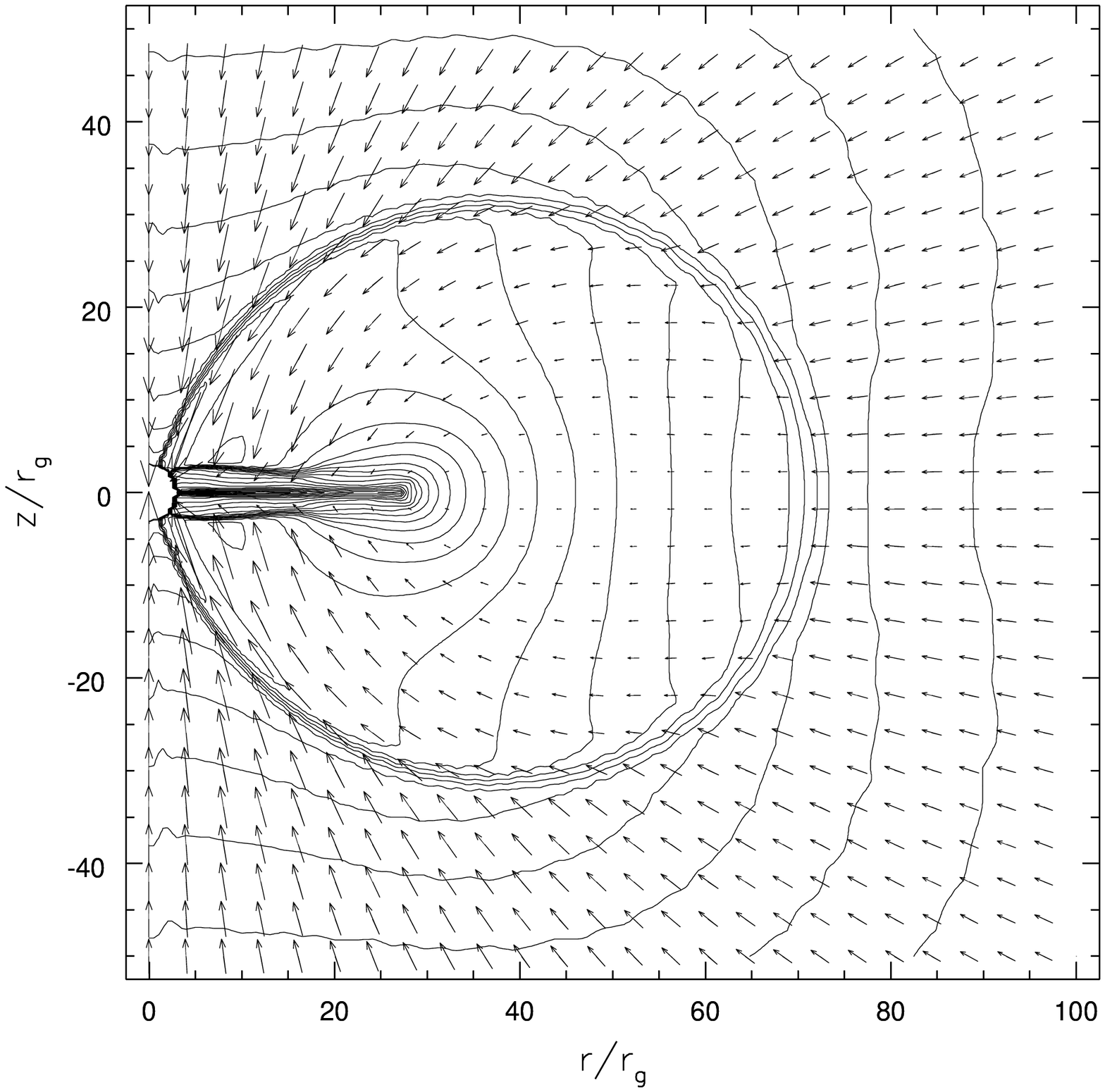}
\caption
%\figcaption
{Low angular momentum accretion onto black hole.
Distributions of density $\rho$
and velocity vector field are shown in the meridional
cross-section for two models with $\dot{M}=0.25\dot{M}_{Edd}$ 
(left) and $0.5\dot{M}_{Edd}$ (right).
Note the different scaling in the axes of the left and right panels.
Vertical axis coincides with the axis of rotation.
Black hole locates in the origin ($0,0$). The contour lines are spaced
with $\Delta\log\rho=0.1$. Cold thin disk, which is not resolved in
our numerical models, locates at the equatorial plane, $z=0$.
Condensation of hot matter to the cold disk takes place inside the radius
$\approx r_d = 30 r_g$ at the equatorial plane.
Accretion flow forms a two-shock structure (shocks are seen
as concentrations of density contours).
The structure is more compact for the higher accretion rate model
(right panel).
In the inner shocks, the supersonic accretion flows,
which locate close to the polar directions, are slowed down before
condensation to the thin disk.
The outer shock originates due to  a centrifugal barrier,
where the infalling matter
is stopped at the equatorial region by the centrifugal force.
\label{fig1}}
\end{figure}

\placefigure{fig1}

Note, that we do not resolve in our calculations the regions close to the thin
disk surface, where the main part of energy of the matter being condensed
is released, and where the Comptonized spectrum is formed.
We are limited in space resolution by the sizes of numerical cells,
but the scale of condensation region is much less than the cell's sizes.
This lack of resolution can be demonstrated by the estimation of
the $y$-parameter, calculated as
the integral $y=\int(kT_e/m_e c^2)d\tau$, 
where $\tau$ is the Thomson optical depth and
the integration is taken in $z$-direction from the equatorial plane till
the outer boundary.
For the larger accretion rate model, $\dot{M}=0.5 \dot{M}_{Edd}$,
the maximum value of $y$ is only of order of $0.1$, whereas the optical
depth in vertical direction does not exceed $0.5$.
The spectrum of the escaping radiation is most probably formed at layers with 
$y\simeq 1$ and $\tau\simeq 1$, 
and its hardness is determined by the temperature of these layers.
To resolve the layers, where the spectrum is formed, one should increase
the resolution of numerical scheme at more than ten times in comparison
with the present one (we currently use $n_r\times n_\theta=150\times 100$).
As an alternative approach we propose to study the vertical structure
of the condensation region using a one-dimension numerical scheme
with the boundary conditions taken from the two-dimension models.
%The latter temperature is consistent with the estimates
%of conditions in plasma in X-ray emitting regions of galactic black
%hole candidates and AGN.

Radiative shocks of discussed type could be thermally unstable
(see Saxton et al. 1998), and their
instabilities may drive quasi-periodic oscillations 
with time scale of order of $10 r_g/c$, which
could explain the observed high frequency variability in the
X-ray flux (Cui et al. 1997).

The outer shock shapes depend mainly on $\dot{M}$ and
vary from an oblate spheroid to a torus (see Fig.~1).
The formation of this shock is
connected to the presence of the centrifugal barrier:  supersonically
moving matter is stopped at $r\ga r_d$ by the centrifugal force.
The possibility of such kind of centrifugal driving shocks in accretion
flows was first pointed out by Fukue 1987. 
In the pre-shock region, between the outer boundary and the outer shock,
the viscous transport of angular momentum 
is not important due to the supersonic accretion velocities.
Slowing down of matter inflow on the outer shock makes
the viscous transport of angular momentum important
in the outer post-shock region.
The angular momentum is effectively transported outward there.
This outward 
transport, from the rapidly rotating inner parts to the slowly rotating
outer ones, is balanced by the inward advection of angular momentum by
the inflowing matter.
This balance, as well as
the efficient cooling of plasma via the Compton mechanism, 
play a significant role in
limiting the size of the outer post-shock region and creation of the
stationary shock structures. 
We found, that in the case
when the Compton cooling is (artificially) switched off, no stationary
solutions of this type are possible.

The dependence of the flow structure on the
parameters $r_d$ and $\dot{M}$ can be used for explanation
of the observed spectral variability of galactic black hole candidates.
For example, the change of spectral state from `hard' to `soft'
in the case of Cygnus X-1
(see Tanaka \& Lewin 1995) can be explained
in terms of variation of $r_d$.
In the hard state (small $r_d$)
most of the accreting matter goes through the
inner shocks at radius $\sim 10 r_g$. In this case a significant part
of radiation comes in hard X-rays.
In the soft state (large $r_d$)
the accretion matter comes to the thin disk
at radii $\ga 100 r_g$. Close to the black hole, accretion through
a thin disk takes place only. 
Obviously, hard X-ray emittion is suppressed in this case.
%For a fixed $r_d$, the accretion flow with larger $\dot{M}$
%produces `softer' radiation from the source,
%due to more efficient Compton cooling, which reduces the electron temperature
%of emitting plasma.
Both parameters $r_d$ and $\dot{M}$ are very sensitive to
conditions of mass exchange between binary companions in the case
of a wind fed accretion (Illarionov \& Sunyaev 1975b).
A small change in the velocity of
the wind of an OB star companion leads to a significant variation
of both $r_d$ and $\dot{M}$ with amplitudes that could explain
observed variability.

\section{Conclusion}

New model of accreting black holes have been proposed, which self-consistently
explains the existence of two-component (hot and cold) plasma in the
vicinity of black holes. The model assumes that at large radial
distances the accretion flow has a quasi-spherical geometry and a low
characteristic angular momentum.
The main parameters of the model are the accretion rate and the amount of
angular momentum carrying by accretion flow.
The model can naturally explain,
without phenomenological assumptions, the origin of hard
X-ray excess observed from the stellar mass black holes
(Galactic black hole candidates), as well as the supermassive black holes
(quasars and AGN).
More hydrodynamical and radiative transfer simulations are needed to
construct detailed spectra of accreting black holes, which will be
compared with observations.

\acknowledgments

This study was supported in part by the
grant from the Royal Swedish Academy of Sciences, RFBR grant 97-02-16975
and NORDITA under the Nordic Project
`Non-linear phenomena in accretion disks around black holes'.
We would like to thank the referee for many helpful suggestions.

%\placefigure{fig2}

\clearpage

\end{document}